\documentclass[twocolumn,showpacs,preprintnumbers,prl]{revtex4}
\usepackage{graphicx}
\usepackage{dcolumn}
\usepackage{bm}
\usepackage{color}

\begin{document}

\title{Shaping single photons and biphotons by inherent losses}
\author{Wei-Ming Su}
\author{Ravikumar Chinnarasu}
\author{Chang-Hau Kuo}
\author{Chih-Sung Chuu}
\email{cschuu@phys.nthu.edu.tw}
\affiliation{Department of Physics, National Tsing Hua University, Hsinchu 30013, Taiwan\\
and Frontier Research Center on Fundamental and Applied Sciences of Matters, National Tsing Hua University, Hsinchu 30013, Taiwan}

\begin{abstract}

Inherent loss is always to be avoided in generating single photons or biphotons, but interestingly it provides opportunities for manipulating the photon wave packet. In this paper we show how inherent loss in parametric down-conversion can be employed to tailor the wave packets of single photons and biphotons. As an example, we propose a scheme to realize a single photon in a single cycle using inherent loss. Our work has potential applications in quantum communication, quantum computation, and quantum interface.  

\end{abstract}


\maketitle

The spatial and temporal extent of a single photon has attracted attention since the early days of quantum mechanics. In 1928 Lawrence and Beams, while studying the photoelectric effect, made the first attempt to determine the longitudinal extent of a photon by ``cutting'' the photon with a fast shutter \cite{Lawrence28}. Today the extent of single photons or biphotons can even be modified by remarkable techniques such as four-wave mixing with electromagnetically induced transparency \cite{Balic05, Du08, Kolchin08}, resonant parametric down-conversion \cite{Ou99, Kuklewicz06, Bao08, Scholz09, Wolfgramm11, Chuu11}, cavity quantum electrodynamics \cite{Kuhn02, Keller04, McKeever04, Thompson06}, coherent pulse-shaping \cite{Peer05}, or chirped quasi-phase-matching~\cite{Harris07, Sensarn10, Carrasco07, Nasr08}. With these new possibilities, the wavefunctions of single photons and biphotons can be controlled to benefit quantum information applications~\cite{Cirac97, Gorshkov07, Novikova07, Zhang12}. 

In this paper we show that the wavepacket of conditional single photons or biphotons in parametric down-conversion (PDC) can be tailored by the \textit{inherent loss} (linear absorption) of the photons, which is always to be avoided in generating single photons or biphotons. As an example, we propose a THz single photon source capable of generating single photons with a temporal length of a single cycle. The electric field within the envelope of the single-cycle single photon completes just one period of oscillation. The inherent loss results from the decay of a phonon polariton due to coupling to acoustic phonons and scattering by crystal defect \cite{Wiederrecht95, Schwarz96, Qiu97}, and depends on the frequency and temperature~\cite{Wu15}.

To describe the signal and idler fields in PDC, non-Langevin theory \cite{Byer68, Kleinman68} may be used if the inherent loss is small or negligible. However, when the inherent loss is large, quantum Langevin theory \cite{Shwartz12, Kolchin07, Harris} is necessary to catch the physics of PDC. This was demonstrated by Shwartz \textit{et. al.}~\cite{Shwartz12} in the x-ray regime, where loss in both signal and idler fields are significant. Here we use quantum Langevin theory to study how inherent loss modifies the photon wavepacket in PDC with one or two lossy quantum fields. The former occurs when the entangled photons have considerably different frequencies and has potential application in wide-band quantum interface~\cite{Ikuta11}, too. We also compare the effect of inherent loss on PDC of the forward-wave type to that of the backward-wave type.

The coupled equations describing the propagation and generation of the signal and idler fields may be formulated in the Heisenberg picture \cite{Shwartz12, Kolchin07, Harris}. Assuming the pump is undepleted and monochromatic at $\omega_p$, we can write the coupled equations as  
\begin{equation}
\left( \textbf{I} \frac{\partial}{\partial z} - \textbf{M} \right) \left[ 
\begin{array}{c}
 a_s (\omega, z) \\
 a_i^{\dagger} (\omega_i, z)
\end{array} \right] = \left[ 
\begin{array}{c}
 \sqrt{\alpha_s} \ f_s (\omega, z) \\
 \sqrt{\alpha_i} \ f_i^{\dagger} (\omega_i, z)
\end{array} \right],
\label{eq:coupled eqs}
\end{equation}
where the signal and idler fields are described by the frequency-domain operators $a_s (\omega, z)$ and $a_i (\omega_i = \omega_p - \omega, z)$, respectively, $\alpha_s$ and $\alpha_i$ are the field absorption coefficients, $f_s (\omega, z)$ and $f_i (\omega_i, z)$ are the Langevin or quantum noise operators, and $\textbf{I}$ is the identity matrix. 

The expression of $\textbf{M}$ depends on the type of parametric interaction. If the interaction is \textit{forward-wave} type, in which the signal and idler photons travel in same direction,
\begin{equation}
\textbf{M} = \left[ 
\begin{array}{cc}
 \alpha_s + i \Delta k (\omega) /2 & - i \kappa \\
 i \kappa^* & \alpha_i - i \Delta k (\omega)/2
\end{array} \right]
\label{eq:M_fw}
\end{equation}
with the coupling constant $\kappa$ and the phase-mismatch $\Delta k (\omega) = k_p (\omega_p) - k_s (\omega) - k_i (\omega_i)$. The signal and idler fields at the output of the generating crystal can then be obtained by solving Eq.~(\ref{eq:coupled eqs}),
\begin{eqnarray}
a_s (\omega, L) = A_1 a_s(\omega, 0) + B_1 a_i^{\dagger} (\omega_i, z_{\rm in})  \nonumber \\
 +\int_0^L E_1 f_s (\omega, z') dz' + \int_0^L F_1 f_i^{\dagger} (\omega_i, z') dz', \label{eq:ABCD_fw1} \\
a_i^{\dagger} (\omega_i, z_{\rm out}) = C_1 a_s(\omega, 0) + D_1 a_i^{\dagger} (\omega_i, z_{\rm in})  \nonumber \\
 +\int_0^L G_1 f_s (\omega, z') dz' + \int_0^L H_1 f_i^{\dagger} (\omega_i, z') dz', \label{eq:ABCD_fw2}
\end{eqnarray}
where $L$ is the crystal length, $z_{\rm out} = L$, $z_{\rm in} = 0$, and
\begin{eqnarray}
\left[ 
\begin{array}{cc}
 A_1 & B_1 \\
 C_1 & D_1
\end{array} \right] &=& e^{-\textbf{M}L}, \label{eq:ABCD_definition1} \\
\left[ 
\begin{array}{cc}
 E_1 & F_1 \\
 G_1 & H_1
\end{array} \right] &=& e^{-\textbf{M}(L-z)} \left[ 
\begin{array}{cc}
 \sqrt{\alpha_s} & 0 \\
 0 & \sqrt{\alpha_i}
\end{array} \right]. \label{eq:ABCD_definition2}
\end{eqnarray}
For simplicity, the variables $\omega$, $\omega_i$, and $z$ are not explicitly shown in $A_1$ to $H_1$. Compared to the lossless PDC ($\alpha_s =\alpha_i = 0$), the output fields of the signal and idler carry additional terms associated with the Langevin noise. The presence of these terms is essential to conserving the commutator relation of the signal and idler field operators. 

If the parametric interaction is \textit{backward-wave} type, in which the idler photons travel in opposite direction of the signal photons and pump, then
\begin{equation}
\textbf{M} = \left[ 
\begin{array}{cc}
 - \alpha_s + i \Delta k (\omega) /2 & - i \kappa \\
 - i \kappa^* & - \alpha_i - i \Delta k (\omega) /2
\end{array} \right]
\label{eq:M'}
\end{equation}
and the phase-mismatch $\Delta k (\omega) = k_p (\omega_p) - k_s (\omega) + k_i (\omega_i)$. The signal and idler fields at the output of the crystal take on similar forms as thos of Eqs.~(\ref{eq:ABCD_fw1}) and (\ref{eq:ABCD_fw2}) but with $z_{\rm out} = 0$, $z_{\rm in} = L$, and $A_1$ to $H_1$ in Eqs.~(\ref{eq:ABCD_fw1}) and (\ref{eq:ABCD_fw2}) replaced by
\begin{eqnarray}
\left[ 
\begin{array}{cc}
 A_2 & B_2 \\
 C_2 & D_2
\end{array} \right] &=& \frac{1}{D_1} \left[ 
\begin{array}{cc}
 A_1 D_1 - B_1 C_1 & B_1 \\
 - C_1 & 1
\end{array} \right], \\
\left[ 
\begin{array}{cc}
 E_2 & F_2 \\
 G_2 & H_2
\end{array} \right] &=& \frac{1}{D_1} \left[ 
\begin{array}{cc}
 E_1 D_1 - B_1 G_1 & F_1 - B_1 H_1 \\
 - G_1 & - H_1
\end{array} \right]. 
\label{eq:ABCD_definition}
\end{eqnarray}
Again, we drop the variables $\omega$, $\omega_i$, and $z$ in $A_2$ to $H_2$ for simplicity.

With the output fields given by Eq.~(\ref{eq:ABCD_fw1}) and (\ref{eq:ABCD_fw2}) and the commutators $[a_j(\omega_1,z_1),a^{\dagger}_k(\omega_2,z_2)]=(2\pi)^{-1}\delta_{jk}\delta(\omega_1-\omega_2)\delta(z_1-z_2)$, $[f_i(\omega_1,z_1),f^{\dagger}_i(\omega_2,z_2)]=(2\pi)^{-1}\delta(\omega_1-\omega_2)\delta(z_1-z_2)$, we can calculate the count rates $R_s = \langle a_s^{\dagger}(\omega)a_s(\omega) \rangle$ and $R_i = \langle a_i^{\dagger}(\omega_i)a_i(\omega_i) \rangle$ of the signal and idler fields, respectively,
\begin{eqnarray}
R_s = \int_{-\infty}^{\infty} S_s (\omega') d\omega', \ \ \ R_i = \int_{-\infty}^{\infty} S_i (\omega') d\omega',
\label{eq:singles rate}
\end{eqnarray}
where 
\begin{eqnarray}
S_s(\omega) &=& \frac{1}{2 \pi} \left( \left| B_k \right|^2 + \int_0^L \left| F_k \right|^2 dz' \right) \label{eq:S1}, \\
S_i(\omega) &=& \frac{1}{2 \pi} \left( \left| C_k \right|^2 + \int_0^L \left| G_k \right|^2 dz' \right) \label{eq:S2},
\end{eqnarray}
are the spectral power density at the signal and idler frequency, respectively, and $k=1$ (forward-wave PDC) or 2 (backward-wave PDC). Note that both $S_s(\omega)$ and $S_i(\omega)$ (and therefore $R_s$ and $R_i$) consist of a parametric and Langevin term, and depend on $\alpha_s$ and $\alpha_i$. The second-order intensity or the Glauber correlation function of the biphotons, $G^{(2)}(\tau) = \langle a_s^{\dagger} (t+\tau) a_i^{\dagger} (t) a_i (t) a_s (t+\tau) \rangle$, can also be obtained as
\begin{equation}
G^{(2)} (\tau) = \frac{1}{(2\pi)^2} \left| \int_{-\infty}^{\infty} \phi (\omega') e^{i \omega' \tau} d \omega' \right|^2 + R_s R_i
\label{eq:G}
\end{equation}
where $\tau = t_i - t_s$ is the time delay between the arrival of the signal and idler photons at the detectors, and 
\begin{equation}
\phi (\omega) = B_k^* D_k + \int_0^L F_k^* H_k dz' 
\end{equation}
is the frequency-domain wavefunction which consists of both parametric and Langevin parts. The waveform of the Glauber correlation function thus depends on the inherent loss of both signal and idler fields.

\begin{figure}[t]
\centering
\includegraphics[width=1 \linewidth]{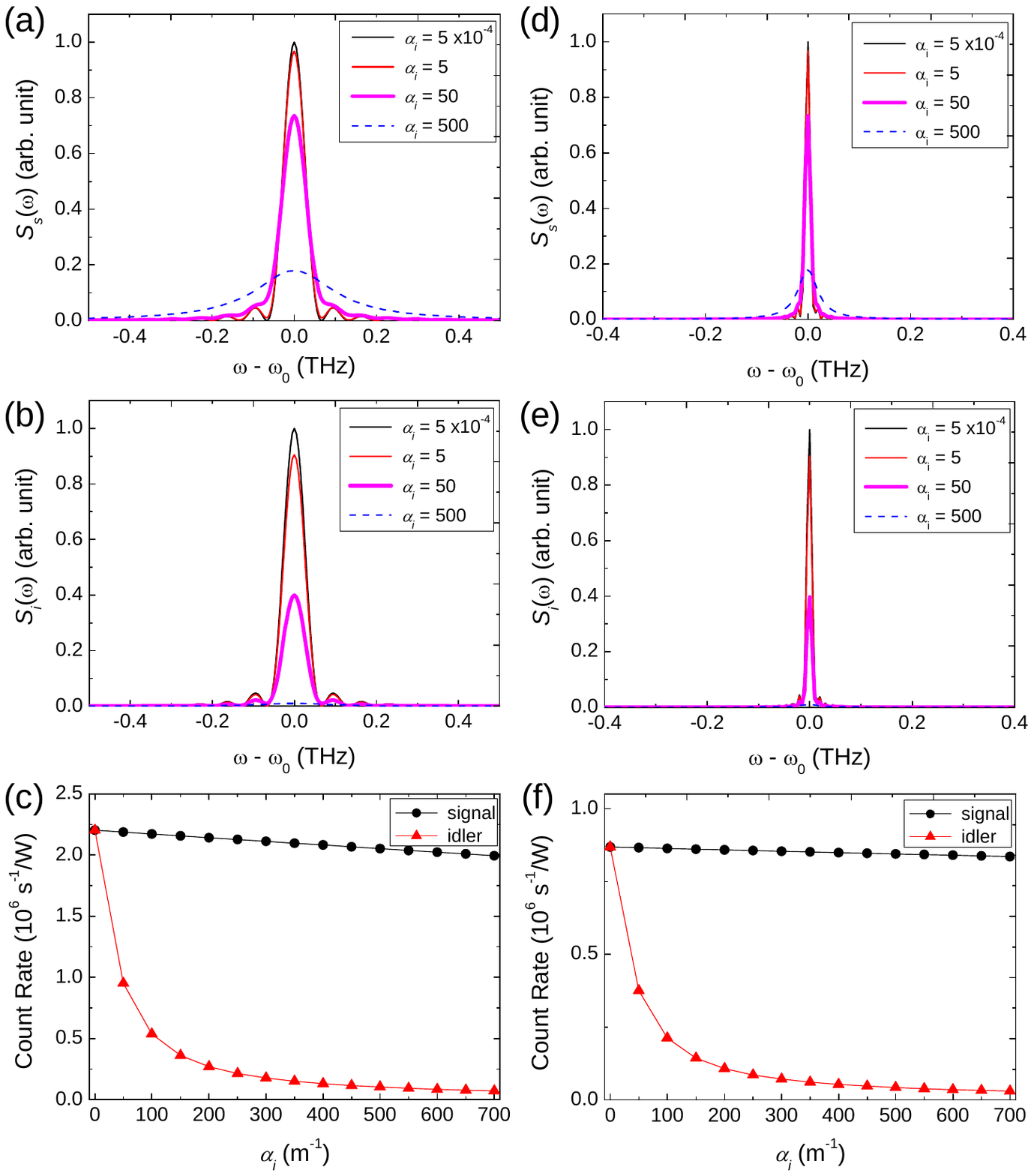}
\caption{\label{fig:3} (color online) In (a), (b), (d), (e): Spectral power density at the signal and idler frequency, $S_s(\omega)$ and $S_i(\omega)$, for $\alpha_i = 0.0005, 5, 50, 500$ m$^{-1}$. In (c), (f): Count rates of the signal and idler photons as a function of absorption coefficient $\alpha_i$. The parametric interaction is forward-wave type in (a)-(c) and backward-wave type in (d)-(f). The crystal length is assumed to be 1 cm in all calculations.}
\end{figure}

We first consider the case in which one quantum field (e.g., the idler field) is lossy but another field is lossless. We are interested in the conditional single photons generated in the PDC, for example, the idler photon conditional on the detection of a signal photon. The wavepacket of the conditional single photon is described by the Glauber correlation function and depends on the inherent loss. This provides a neat way of manipulating the wavefunction of the single photon through inherent loss. To gain insights into how the inherent loss of the idler field affects the properties of the signal and idler fields as well as their quantum correlation, we assume a 1-cm-long generating crystal with $\alpha_s \approx 0$ and $\alpha_i$ varying from 0.0005 to 500 m$^{-1}$. In Fig.~\ref{fig:3}~(a) and Fig.~\ref{fig:3}~(b) (forward-wave PDC) and Fig.~\ref{fig:3} (d) and Fig.~\ref{fig:3}~(e) (backward-wave PDC), we plot the spectral power density at the signal and idler frequencies for various $\alpha_i$. When the inherent loss is negligible, we see the familiar sinc$^2$ shape as for a lossless PDC. However, as the inherent loss increases, both $S_s(\omega)$ and $S_i(\omega)$ differ from the sinc$^2$ shape and turn into a Lorentzian-like function with larger bandwidths. This is also accompanied by the decrease of the signal and idler count rates, $R_s$ and $R_i$, as shown in Fig.~\ref{fig:3} (c) (forward-wave PDC) and Fig.~\ref{fig:3}~(f) (backward-wave PDC). Although both $R_s$ (black dots) and $R_i$ (red triangles) decrease with $\alpha_i$, $R_s$ changes more gradually than $R_i$. It is because the parametric contribution of $R_s$, which is associated with the generation of idler field and decreases with $\alpha_i$, is compensated for by the Langevin contribution from quantum noise that increases with $\alpha_i$.

\begin{figure}[t]
\centering
\includegraphics[width=1 \linewidth]{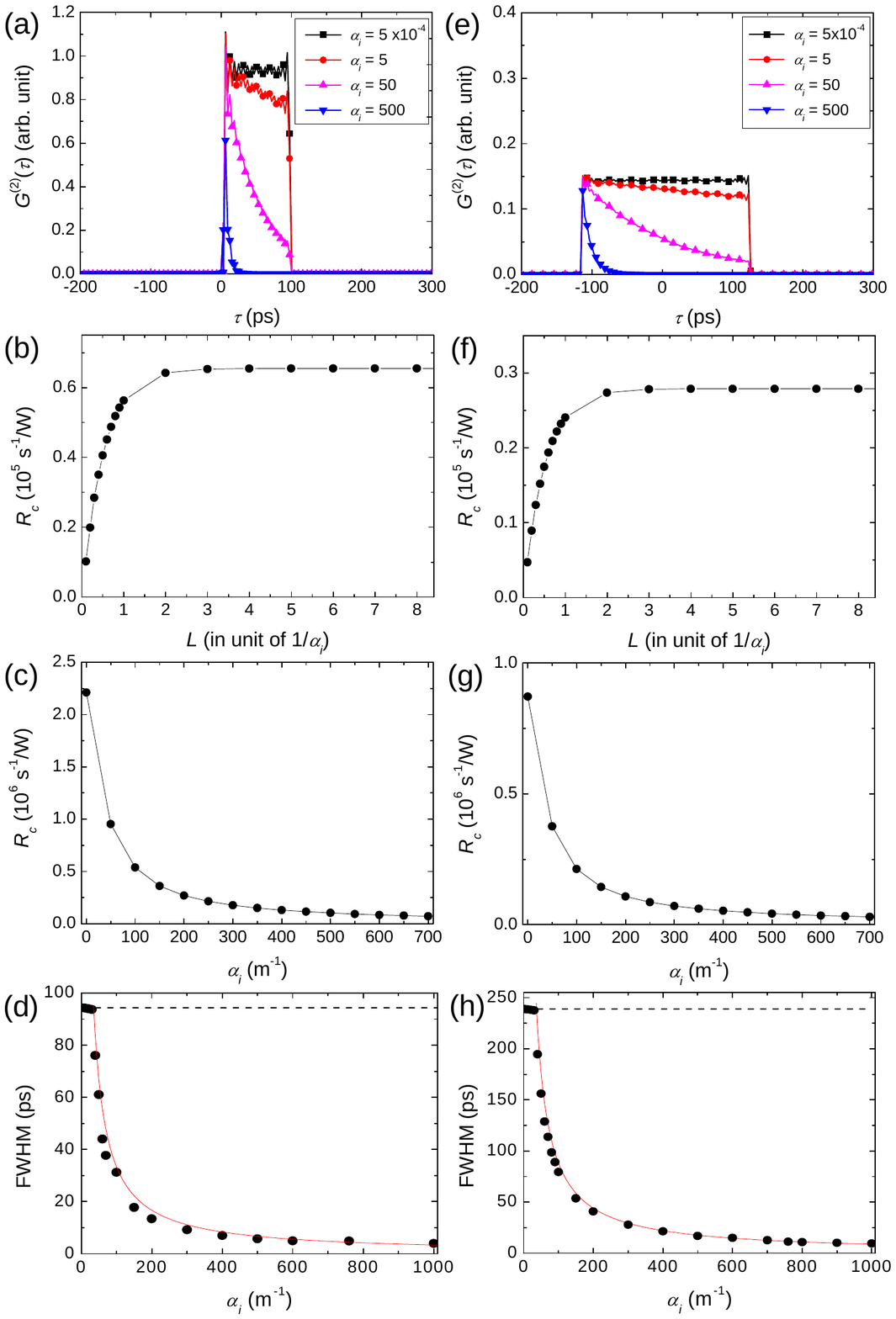}
\caption{\label{fig:4} (color online) (a), (e) Glauber correlation functions for $\alpha_i = 0.0005, 5, 50, 500$ m$^{-1}$. (b), (f) Coincidence rate (per W of pump power) as a function of crystal length (in unit of absorption length). (c), (g) Coincidence rate as a function of absorption coefficient. (d), (h) Full width at half maximum of $G^{(2)}(\tau)$ as a function of absorption coefficient. The parametric interaction is forward-wave type in (a)-(d), and backward-wave type in (e)-(h). The crystal length is assumed to be 1 cm in all calculations.}
\end{figure}

The Glauber correlation functions $G^{(2)}(\tau)$ are illustrated in Fig.~\ref{fig:4} (a) (forward-wave PDC) and Fig.~\ref{fig:4}~(e) (backward-wave PDC) for various inherent loss. As $\alpha_i$ goes to 0, $G^{(2)}(\tau)$ shows the familiar square shape with a correlation time $L/V_-$ in the forward-wave PDC and $L/V_+$ in the backward-wave PDC, where $V_{\mp} = |1/V_s {\mp} 1/V_i|^{-1}$ is the relative group velocity. The co-propagation of the signal and idler photons in the forward-wave PDC results in a shorter correlation time and a quantum correlation only in the positive time delay. As $\alpha_i$ increases, the absorption of the idler field introduces a decay in correlation that varies with the time delay. Nonetheless, at low inherent loss ($\alpha_i < 1/L$), the square shape is still the dominant feature of $G^{(2)}(\tau)$ and the correlation time is determined by the crystal length. At high inherent loss ($\alpha_i > 1/L$), the decay overtakes the square shape and the correlation time is determined by the absorption length. This is because the idler photons, once generated, will be absorbed after traveling a distance of $\alpha_i^{-1}$. Hence, only the idler photons generated within $\alpha_i^{-1}$ to the crystal end will contribute to the correlation. This can also be seen in the coincidence rate $R_c = \int_{-\infty}^{\infty} G^{(2)} (\tau') d \tau'$ of the signal and idler photons in Figs.~\ref{fig:4}~(b) and (f), where $R_c$ approaches a constant value for $\alpha_i L > 1$. In Figs.~\ref{fig:4}~(c) and (g), we also show the coincidence rate as a function of $\alpha_i$. The absorption of the idler field results in a nearly exponential decay of coincidence rate with a $1/e$ decay constant of about $1/L$, a feature that can also be found in the idler count rate.

The FWHM of $G^{(2)}(\tau)$ determines the temporal length of the conditional single photon. As shown in Fig.~\ref{fig:4} (d) (forward-wave PDC) and Fig.~\ref{fig:4} (h) (backward-wave PDC), the FWHM (dots) at low inherent loss is approximately $L/V_-$ (dashed) in the forward-wave PDC and $L/V_+$ (dashed) in the backward-wave PDC. At high inherent loss, the FWHM  to a good approximation is $1/(c\alpha_i)$ (red curve) in the forward-wave PDC and $(V_+/V_-)/(c\alpha_i)$ (red curve) in the backward-wave PDC. The FWHM in the forward-wave PDC is thus always a factor of $V_+/V_-$ shorter than that in the backward-wave PDC. Moreover, the FWHM at high inherent loss can be significantly shorter than the FWHM at low inherent loss, which greatly raises the possibility of generating ultra-short single photons at high inherent loss.

\begin{figure}[t]
\centering
\includegraphics[width=1 \linewidth]{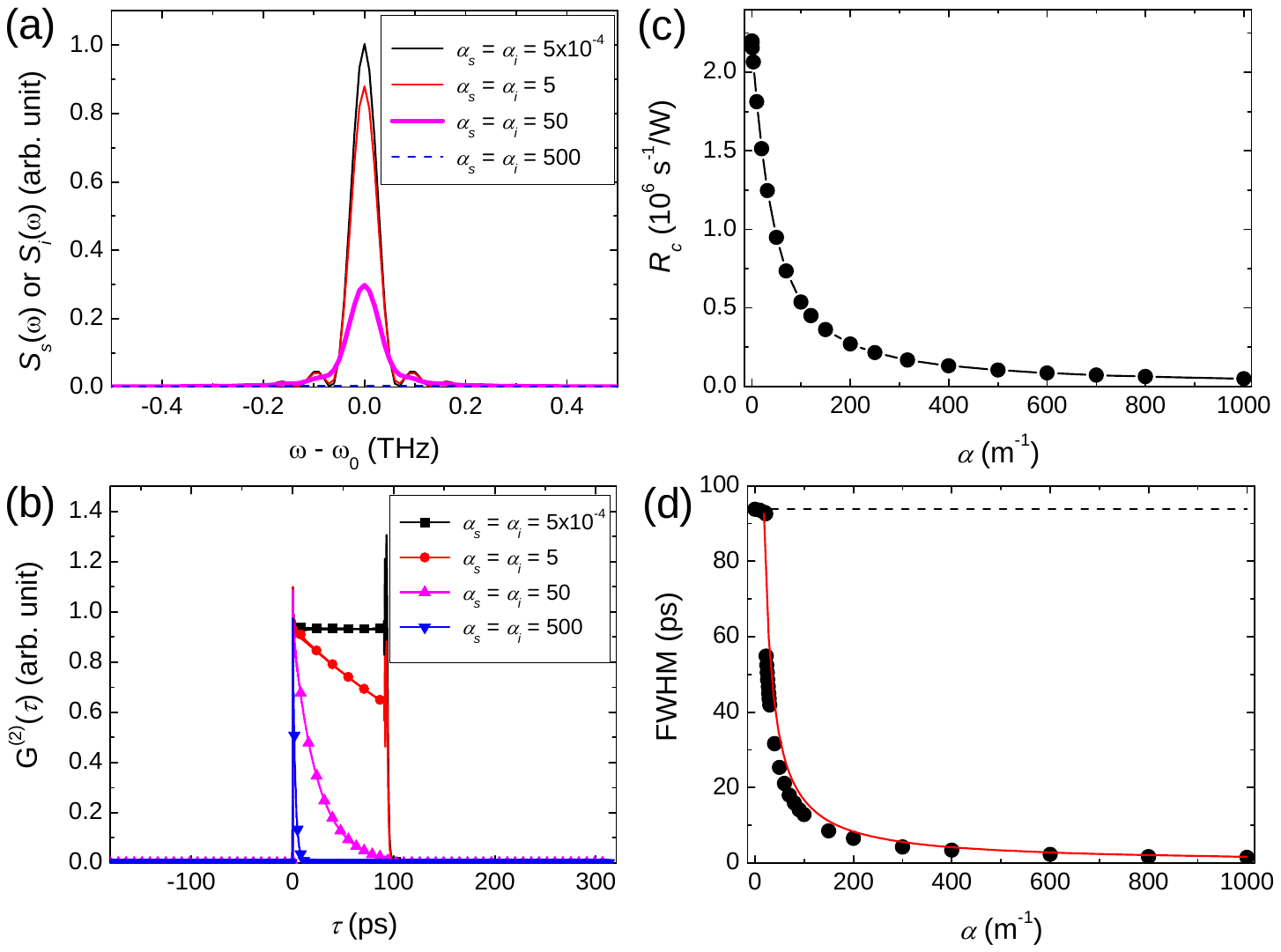}
\caption{\label{fig:6} (color online) (a) Spectral power density at the signal and idler frequency for $\alpha = 0.0005, 5, 50, 500$~m$^{-1}$. (b) Glauber correlation functions $G^{(2)}(\tau)$ for $\alpha = 0.0005, 5, 50, 500$ m$^{-1}$. (c) Coincidence rate as a function of absorption coefficient. (d) Full width at half maximum of $G^{(2)}(\tau)$ as a function of absorption coefficient. The crystal length is assumed to be 1 cm in all calculations.}
\end{figure}

We now turn to the case when both signal and idler fields are lossy. We consider forward-wave interaction and assume that the frequencies of the photon pair are degenerate or nearly degenerate so that $\alpha_s = \alpha_i = \alpha$. The calculated spectral power density $S_s(\omega)$ and $S_i(\omega)$, Glauber correlation function $G^{(2)}(\tau)$, coincidence rate $R_c$, and FWHM are illustrated in Fig.~\ref{fig:6}. The addition of $\alpha_s$ changes all quantities except $S_i(\omega)$: $S_s(\omega)$ [Fig.~\ref{fig:6}(a)] is now identical to $S_i(\omega)$ due to the equal amount of loss, which shows the same dependence on $\alpha$ as $S_i(\omega)$ in PDC with a single lossy field. As $\alpha$ increases, $G^{(2)}(\tau)$ [Fig.~\ref{fig:6}(b)] changes more rapidly from a parametric dominated (square) shape to Langevin dominated (exponential decay) shape. The coincidence rate [dots in Fig.~\ref{fig:6}(c)] also decays more rapidly with a $1/e$ decay constant of $1/(2L)$ instead of $1/L$ as in PDC with a single lossy field. In addition, the FWHM of $G^{(2)}(\tau)$ [dots in Fig.~\ref{fig:6}(d)] is approximately $1/(2c\alpha)$ (red curve) instead of $1/(c\alpha)$ at high inherent loss. The presence of inherent loss in the signal field is thus as if the idler field in PDC of single lossy field experiences an inherent loss of $2\alpha$.

To see how inherent loss may be used to realize a single photon in a single cycle, we consider the forward-wave PDC with signal and idler photons generated at the infrared and THz frequencies, respectively. Photon pairs of this sort can be generated by pumping a periodically poled lithium niobate (LN) crystal with a cw 1.064-$\mu$m laser. The absorption coefficients of LN crystal at the infrared and THz frequencies are about $10^{-3} \ {\rm cm}^{-1}$ \cite{Gettemy88} and $10^2 \ {\rm cm}^{-1}$ \cite{Wu15}, respectively. Taking the crystal length to be 1 cm, we can therefore neglect the loss of the signal photons.

\begin{figure}[t]
\centering
\includegraphics[width=0.95 \linewidth]{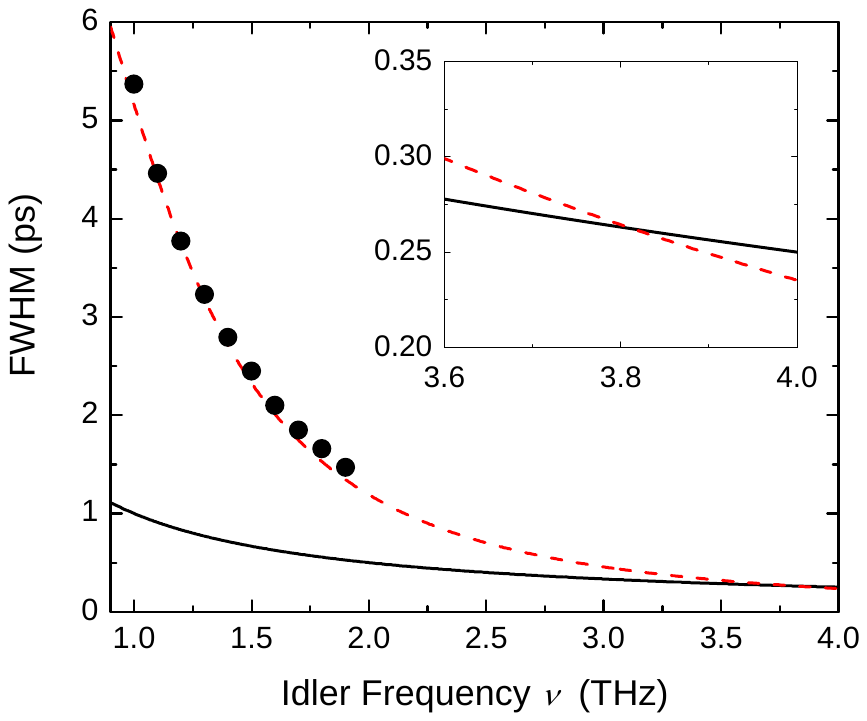}
\caption{\label{fig:5} (color online) FWHM of the conditional idler photon as a function of frequency $\nu$. The dots are calculated by Eq.~(\ref{eq:G}). The dashed line is calculated by $1/c\alpha_i(\nu)$. The solid line represents the single-cycle limit, ${\rm FWHM} = 1/\nu$. Single-cycle single photons are generated at 3.82 THz (inset).}
\end{figure}

A single-cycle idler photon can be generated if the FWHM of its wavepacket equals the inverse of its frequency, namely ${\rm FWHM} = 1/\nu$. In Fig.~\ref{fig:5} we show the FWHM (dots) calculated by Eq.~(\ref{eq:G}) as a function of the idler frequency between 1 and 2 THz, where the Sellmeier equation and absorption coefficient in the THz regime are both available \cite{Wu15, Kiessling13}. The FWHM decreases with the idler frequency due to the larger inherent loss at higher idler frequency. To obtain the FWHM beyond 2 THz, we calculate the FWHM by $1/c\alpha_i(\nu)$ with the approximate form of absorption coefficient $\alpha_i (\nu) = (16.84 - 27.43 \ \nu + 23.51 \ \nu^2) \times 50$ \cite{Wu15}. The result is in good agreement with that obtained by Eq.~(\ref{eq:G}) (dots) between 1 -- 2 THz. The obtained FWHM (dashed line) is then compared to the ${\rm FWHM} = 1/\nu$ (solid line) required for achieving the single-cycle limit. As shown in the inset of Fig.~\ref{fig:5}, the two curves intersect at 3.82 THz where the idler photon is generated in a single cycle.

In summary, we have shown that the inherent loss in PDC provides opportunities for manipulating the wavepacket of single photons or biphotons. This is illustrated in a proposed parametric down-converter which generates conditional THz single photons in a single cycle. The THz single photon, of which the detection has been made possible by the recent development of the photon-counting technique~\cite{Komiyama11}, has potential application in quantum computation \cite{Sherwin99} or communication in particular atmospheric condition \cite{Nagatsuma16}. Single-cycle single photon can also find application in high-bandwidth quantum communication \cite{Drummond14}. Although THz single-cycle single photons are studied in the example, the proposed method works for any frequency regime in principle provided that nonlinear materials with appropriate inherent loss can be found. Compared to the technique of chirped quasi-phase-matching~\cite{Harris07, Sensarn10, Carrasco07, Nasr08}, which can also generate ultrashort photons, our method does not require sophisticated fabrication of a parametric medium although it comes with the cost of reduced photon rate. Single-cycle single photons may also be realized using a lossless medium of length on the order of the photon wavelength. However, it will be very challenging, if not impossible, to generate single-cycle single photons in the infrared and visible regimes, because crystal lengths of a few micrometers or shorter are necessary. Our method, which can be implemented with any medium length longer than the absorption length, is thus highly beneficial at these wavelengths.

The authors thank S. E. Harris for helpful discussion. This work was supported by the Ministry of Science and Technology, Taiwan (MOST 103-2112-M-007-015-MY3). 


\end{document}